\documentclass[12pt]{iopart}
\usepackage{iopams}
\usepackage{dsfont}
\usepackage[dvips]{graphicx}
\newcommand{\half}{\mbox{$\textstyle \frac{1}{2}$}}

\newcommand{\re}{\mbox{$\rm e$}}
\newcommand{\ri}{\mbox{$\rm i$}}

\begin{document}
\newtheorem{theorem}{Theorem}
\newtheorem{corollary}{Corollary}
\newtheorem{prop}{Proposition}

\title[Constrained quantum motion]
{Symplectic approach to quantum constraints}

\author[D~C~Brody, A~C~T~Gustavsson, and
L~P~Hughston]{Dorje~C.~Brody${}^1$, Anna~C.~T.~Gustavsson${}^2$, and
Lane~P.~Hughston${}^1$}

\address{${}^1$Department of Mathematics, Imperial College London,
London SW7 2AZ, UK}

\address{${}^2$Blackett Laboratory,
Imperial College London, London SW7 2AZ, UK}



\begin{abstract}
A general prescription for the treatment of constrained quantum
motion is outlined. We consider in particular constraints defined by
algebraic submanifolds of the quantum state space. The resulting
formalism is applied to obtain solutions to the constrained dynamics
of systems of multiple spin-$\half$ particles. When the motion is
constrained to a certain product space containing all of the energy
eigenstates, the dynamics thus obtained are quasi-unitary in the
sense that the equations of motion take a form identical to that of
unitary motion, but with different boundary conditions. When the
constrained subspace is a product space of disentangled states, the
associated motion is more intricate. Nevertheless, the equations of
motion satisfied by the dynamical variables are obtained in closed
form.
\end{abstract}

\submitto{\JPA}
%
%
\vspace{0.1in}

Practical implementations of quantum algorithms in quantum
information technology can be hampered by constraints. The
formulation of a tractable theory of the properties of constrained
unitary motions is therefore of interest and importance. This paper
provides a framework for dealing with certain classes of constrained
quantum motion, for which differential equations governing the
constrained dynamics can be obtained explicitly. The results are
applied to constrain unitary motions to algebraic subspaces of
quantum state spaces.

The structure and the findings of the paper can be summarised as
follows. We begin by reviewing the Hamiltonian formulation of
standard quantum mechanics. This highly effective way of looking at
quantum mechanics has been investigated by a number of authors (see
\cite{strocchi,cantoni,Kibble,Weinberg,cirelli,gibbons,Hughston,
ashtekar,gqm} and references cited therein). We demonstrate the
existence of a particularly simple choice of canonically conjugate
coordinates $(\{q_i\},\{p_i\})$ for the space of pure quantum states
with the property that the Schr\"odinger equation can be written in
Hamiltonian form:
\begin{eqnarray}
\dot{q}_i=\frac{\partial H(q,p)}{\partial p_i} \quad {\rm and} \quad
\dot{p}_i=-\frac{\partial H(q,p)}{\partial q_i}. \label{eq:1}
\end{eqnarray}
The coordinates $(\{q_i\},\{p_i\})$ appearing here are defined by
the expansion coefficients of the normalised state vector
$|\psi\rangle$ of a non-degenerate $n$-level system in terms of the
energy eigenstates $\{|E_i\rangle\}_{i=1,2,\ldots,n}$ according to
the scheme
\begin{eqnarray}
|\psi\rangle = \sum_{i=1}^{n-1} \surd{p_i}\re^{-{\rm i}q_i}
|E_i\rangle + \left(1-\sum_{i=1}^{n-1}p_i\right)^{\frac{1}{2}}
|E_n\rangle, \label{eq:2}
\end{eqnarray}
where the function
\begin{eqnarray}
H(q,p)=\frac{\langle \psi|{\hat H}|\psi\rangle}{\langle
\psi|\psi\rangle}
\end{eqnarray}
is given by the expectation of the Hamiltonian operator in the state
defined by (\ref{eq:2}). We choose the overall phase such that the
coefficient of the $n$th energy eigenstate $|E_n\rangle$ is real.
Since in the energy basis we have
\begin{equation}
{\hat H}= \sum_{i=1}^{n} E_i |E_i\rangle\langle E_i|,
\end{equation}
it follows that the Hamiltonian function is given by
\begin{eqnarray}
H(q,p) = E_n+\sum_{i=1}^{n-1} \omega_i p_i, \label{eq:3}
\end{eqnarray}
where $\omega_i=E_i-E_n$. Note that $H(q,p)$ is independent of
$\{q_i\}$ and is linear in $\{p_i\}$. Substitution of (\ref{eq:3})
in (\ref{eq:1}) shows that the solution to the Schr\"odinger
equation
\begin{equation}
\ri\frac{\partial}{\partial t}|\psi_t\rangle={\hat H}|\psi_t\rangle
\label{eq:5-SE}
\end{equation}
is given
in terms of the canonical coordinates by
\begin{eqnarray}
q_i(t)=q_i(0)+\omega_i t \quad {\rm and} \quad p_i(t)=p_i(0),
\label{eq:4}
\end{eqnarray}
which should also be evident from (\ref{eq:2}). For simplicity we
shall consider Hamiltonians having nondegenerate eigenvalues,
although the formalism can be applied to degenerate systems by use
of the L\"uders projection postulate \cite{Luders}. Specifically,
the state $|\psi\rangle$ is still expressible in the form
(\ref{eq:2}), but we make the replacement $n\to d$, where $d<n$ is
the number of distinct energy eigenvalues, and $|E_i\rangle =
\hat{\Pi}_i|\psi\rangle$, where $\hat{\Pi}_i$ is the projection
operator onto the eigenspace associated with the eigenvalue $E_i$,
given by
\begin{eqnarray}
\hat{\Pi}_i= \sum_{j=1}^{d_i} |E_i,j\rangle\langle E_i,j|.
\end{eqnarray}
Here $d_i$ is the dimension of the Hilbert subspace associated with
the eigenvalue $E_i$, and $|E_i,j\rangle$ $(j=1,\ldots,d_i)$
constitute an orthonormal basis for that subspace. The Hamiltonian
for a general system is hence given by
\begin{eqnarray}
\hat{H}=\sum_{i=1}^d E_i \hat{\Pi}_i,
\end{eqnarray}
and one sees that the form of (\ref{eq:3}) remains unchanged, except
that $n$ is replaced by $d$, and $H(q,p)$ is independent of the
remaining phase space degrees of freedom.

The objective of this paper is to introduce a framework for treating
certain classes of constrained unitary motion. Our approach is
aligned closely with that of Dirac's theory of constraints in
classical mechanics \cite{Dirac-50,Dirac-58}. The idea that Dirac's
methodology might be applied to investigate constrained quantum
motion was proposed recently by Buri\'c~\cite{Buric} to determine
the dynamics of a pair of spin-$\half$ particles constrained to a
special surface of product states containing all of the energy
eigenstates. An alternative approach to quantum constraints is
considered in \cite{Corichi}.

An elementary way of enforcing the constraints is to introduce
Lagrange multipliers. In some circumstances the Lagrange multipliers
can be determined explicitly and eliminated from the equations of
motion. In this paper we shall be considering such cases. Several
examples are investigated, including one for which the motion of a
pair of spin-$\half$ particles is constrained to the hypersurface of
disentangled states.

Constrained unitary motions are in general nonunitary, and
correspond to nonlinear evolutions. However, unlike the general
nonlinear dynamics of the Mielnik-Kibble-Weinberg (MKW)
framework~\cite{Kibble,Weinberg,Mielnik}, the nonlinearities
resulting from the class of constraints considered in the present
investigation are of a milder form. That is, while in the MKW theory
one considers a general Hamiltonian $H(q,p)$ that is distinct from
(\ref{eq:3}), in the present context the `linear' Hamiltonian
(\ref{eq:3}) remains unchanged, and the nonlinearity arises from a
modification of the symplectic structure, or equivalently on account
of the nonlinearity of the constraint surface.

In the first example we consider a system of two spin-$\frac{1}{2}$
particles. The constraint surface is defined by the product space of
a pair of two-dimensional Hilbert spaces, and we assume that the
product space contains the energy eigenstates. An initial state that
lies on the constraint surface is thus obliged to remain on this
surface. We shall obtain the trajectories of the unitary evolutions
subject to this constraint. The analysis can be extended and applied
to $n$ spin-$\half$ particles constrained to a special product space
containing the energy eigenstates. We find that this constraint is
satisfied under unitary evolution without constraint if the spectrum
of the trace-free part of the Hamiltonian takes the form
$\{E_i\}=\{e_1,e_2,\cdots, e_{n/2},-e_{n/2}, -e_{n/2-1}, \cdots,
-e_{1}\}$, where $n$ is the number of eigenstates of the system. For
a generic Hamiltonian we derive and solve the constrained equations
of motion explicitly in the case of two and three spin-$\frac{1}{2}$
particle systems. Surprisingly, the resulting dynamics turn out to
be quasi-unitary in the sense that the amplitudes $\{p_i\}$ remain
constant while the relative phases $\{q_i\}$ evolve linearly. This
property appears to be generic even for systems with more particles.
We then consider the motion of two spin-$\frac{1}{2}$ particles
constrained to remain on the quadric corresponding to the subspace
of disentangled states.

Let us now begin with a summary of the Hamiltonian formulation of
quantum mechanics. We consider a complex Hilbert space ${\mathcal
H}^n$ of dimension $n$, a typical element of which is denoted
$|\psi\rangle$. If ${\hat F}$ represents an observable, its
expectation with respect to $|\psi\rangle$ is $F=\langle\psi|{\hat
F} |\psi\rangle/ \langle\psi|\psi\rangle$, which is invariant under
the transformation $|\psi\rangle\to\lambda|\psi\rangle$,
$\lambda\in{\mathds C}-\{0\}$. The vector $|\psi\rangle$ thus
carries a redundant complex degree of freedom. We therefore
construct the space of rays through the origin of ${\mathcal H}^n$
by the identification $|\psi\rangle\sim \lambda |\psi\rangle$. The
result is the projective Hilbert space ${\mathcal P}^{n-1}$ of
dimension $n-1$. We can view ${\mathcal P}^{n-1}$ as a real
even-dimensional manifold $\Gamma$, and let $\{x^a\}_{a=1,2,\ldots,
2n-2}$ denote a typical point of $\Gamma$. One distinguishing
feature of $\Gamma$ is that it is equipped with a symplectic
structure $\Omega^{ab}(=-\Omega^{ba})$ such that the Schr\"odinger
equation (\ref{eq:5-SE}) can be expressed in the Hamiltonian form
\begin{eqnarray}
\dot{x}^a = \Omega^{ab}\nabla_b H. \label{eq:5}
\end{eqnarray}
Here the function $H(x)=\langle\psi(x)|{\hat H}|\psi(x)\rangle/
\langle\psi(x)|\psi(x)\rangle$ denotes the expectation of the
operator ${\hat H}$ in the pure state $|\psi(x)\rangle$
corresponding to the point $x\in\Gamma$. One can regard $\Gamma$ as
a bona fide quantum analogue of the phase space in classical
mechanics.

Now suppose that we have a family of constraints on the motion of
the system in $\Gamma$ expressed in the form
\begin{eqnarray}
\Phi^{\alpha}(x) = 0, \label{eq:6}
\end{eqnarray}
where $\alpha=1,2,\ldots,N$. There are two different types of
constraints that arise naturally in the quantum context,
corresponding to what one might call `algebraic' and `real'
constraints. In the algebraic case the motion is confined to an
algebraic submanifold (or possibly a complex algebraic subvariety)
of the original quantum state space ${\mathcal P}^{n-1}$. As a
consequence, the constraint submanifold is of even real
dimension---for which it follows that the number of constraints $N$
is \textit{even} in this case. This paper is primarily concerned
with the algebraic case. Typical examples would include the
situations where the constraint manifold ${\mathfrak M}$ was an
algebraic curve in ${\mathcal P}^2$ (such as a conic or a elliptic
cubic curve), or an algebraic curve in ${\mathcal P}^3$ (such as a
twisted cubic curve or an elliptic quartic curve), or an algebraic
surface in ${\mathcal P}^3$ (such as a quadric surface, or a cubic
surface). The equations given by (\ref{eq:6}) then define
${\mathfrak M}$ locally.

The other situation that is natural to consider in quantum theory is
the case where the constraints are of the form
\begin{eqnarray}
\Phi^\alpha(x) = \frac{\langle\psi(x)|{\hat F}^\alpha
|\psi(x)\rangle}{\langle\psi(x)|\psi(x)\rangle} - f^\alpha,
\end{eqnarray}
where $\{{\hat F}^\alpha\}_{\alpha=1,\ldots,N}$ denotes a collection
of observables, $\{f^\alpha\}_{\alpha=1,\ldots,N}$ is a set of real
numbers, and $N$ need not be even. We shall consider the `real' case
elsewhere.

In the algebraic case, the constraints can be enforced by the
introduction of Lagrange multipliers
$\{\lambda_\alpha\}_{\alpha=1,2,\ldots,N}$. Using the usual
summation convention, the constrained equations of motion are
\begin{equation}
\dot{x}^a = \Omega^{ab} \nabla_b H + \lambda_{\alpha} \Omega^{ab}
\nabla_b \Phi^{\alpha}. \label{eq:7}
\end{equation}
To determine the Lagrange multipliers we analyse the relation
$\dot{\Phi}^\alpha(x)=0$. From the chain rule we have
$\dot{\Phi}^{\alpha}=\dot{x}^a \nabla_a \Phi^{\alpha} = 0$.
Substituting (\ref{eq:7}) in here, we find
\begin{equation}
\Omega^{ab} \nabla_a \Phi^{\alpha} \nabla_b H + \lambda_{\beta}
\Omega^{ab} \nabla_a \Phi^{\alpha}\nabla_b \Phi^{\beta} = 0.
\label{eq:9}
\end{equation}
To solve (\ref{eq:9}) for $\lambda_{\alpha}$ let us define
\begin{equation}
\omega^{\alpha\beta} = \Omega^{ab} \nabla_a \Phi^{\alpha} \nabla_b
\Phi^{\beta} . \label{eq:10}
\end{equation}
In the case of real constraints, for which $\{\Phi^\alpha\}$
corresponds to a family of observables, $\omega^{\alpha\beta}$ is
the commutator of the observables $\Phi^\alpha$ and $\Phi^\beta$. We
note that since $\Omega^{ab}= -\Omega^{ba}$ we have
$\omega^{\alpha\beta} =-\omega^{\beta\alpha}$.

If the matrix $\omega^{\alpha\beta}$ is nonsingular, then we can
invert it. In that case, writing $\omega_{\alpha\beta}$ for the
inverse of $\omega^{\alpha\beta}$ so $\omega^{\alpha\beta}
\omega_{\beta\gamma} = \delta^{\alpha}_{\phantom{\alpha}\gamma}$, we
can solve (\ref{eq:9}) for $\{\lambda_\alpha\}$ to obtain
\begin{equation}
\lambda_{\alpha} = \omega_{\beta\alpha} \Omega^{ab} \nabla_a
\Phi^{\beta} \nabla_b H. \label{eq:11}
\end{equation}
Substituting this in the right side of (\ref{eq:7}) yields
\begin{equation}
\dot{x}^a = \Omega^{ab} \nabla_b H + \omega_{\beta\alpha}
\Omega^{cd} \nabla_c \Phi^{\beta} \nabla_d H \Omega^{ab} \nabla_b
\Phi^{\alpha}. \label{eq:12}
\end{equation}
This can be simplified further by rearrangement of indices, after
which we deduce that
\begin{equation}
\dot{x}^a = {\tilde\Omega}^{ab} \nabla_b H, \label{eq:13}
\end{equation}
where ${\tilde\Omega}^{ab}=\Omega^{ab} + \Lambda^{ab}$ and
\begin{equation}
\Lambda^{ab} = \Omega^{ac} \Omega^{bd} \omega_{\gamma \delta}
\nabla_c \Phi^{\gamma} \nabla_d \Phi^{\delta}. \label{eq:14}
\end{equation}
An important point to note is that $\Lambda^{ab}$ is by construction
antisymmetric. Therefore, ${\tilde\Omega}^{ab}$ defines a modified
symplectic structure. The constrained equation of motion
(\ref{eq:13}) thus takes on a form identical to (\ref{eq:5}), with
the same Hamiltonian, but with the modified symplectic structure.

The modified symplectic structure can be interpreted as playing the
role of an induced symplectic structure on the constraint surface
$\Phi=0$. To see this we transvect $\tilde{\Omega}^{ab}$ with the
vector $\nabla_a \Phi^{\alpha}$ normal to the constraint surface to
obtain
\begin{eqnarray}
\tilde{\Omega}^{ab}\nabla_b \Phi^{\alpha} = \Omega^{ab}
\nabla_b\Phi^{\alpha} + \Omega^{ac}\Omega^{bd}\omega_{\gamma\delta}
\nabla_c\Phi^{\gamma} \nabla_d\Phi^{\delta}\nabla_b\Phi^{\alpha}.
\end{eqnarray}
Using the antisymmetry of $\Omega^{ab}$ and the definition
(\ref{eq:10}) we find
\begin{eqnarray}
\Omega^{bd} \nabla_d \Phi^{\delta} \nabla_b \Phi^{\alpha} =
-\omega^{\delta\alpha}.
\end{eqnarray}
Hence from
$\omega_{\gamma\delta}\omega^{\delta\alpha}=\delta_{\gamma}^{\alpha}$
we deduce that
\begin{eqnarray}
\tilde{\Omega}^{ab}\nabla_b \Phi^{\alpha} = 0
\end{eqnarray}
for all $\alpha$. Therefore, $\tilde{\Omega}^{ab}$ annihilates all
vectors normal to the constraint surface, and hence induces a
symplectic structure on the constraint surface.

Our procedure for dealing with a constrained unitary motion can be
summarised as follows: (i) find a suitable choice of $2n-2$ real
coordinates for representing the generic pure state $|\psi\rangle$;
(ii) calculate the symplectic structure $\Omega^{ab}$ in that
coordinate system so that the unitary evolution is represented in
the Hamiltonian form (\ref{eq:5}); (iii) express the constraints
(\ref{eq:6}) in terms of the given choice of coordinates; (iv)
assuming that the constraints are such that the matrix
$\omega^{\alpha\beta}$ of (\ref{eq:10}) is invertible, calculate
$\Lambda^{ab}$ according to (\ref{eq:14}) and substitute the result
into (\ref{eq:13}). In this way, dynamical equations for constrained
unitary motion can be obtained, and one is left with the problem of
solving a system of coupled differential equations.

As we have indicated, there is a particular choice of coordinates on
the space of pure states for which the analysis can be simplified in
the form defined in (\ref{eq:2}), which might appropriately be
called an `action-angle' parametrisation (cf. \cite{oh}). In terms
of these coordinates $\Omega^{ab}$ is given by
\begin{eqnarray}
\Omega^{ab}=\left( \begin{array}{cc} {\mathds O} & {\mathds 1} \\
-{\mathds 1} & {\mathds O} \end{array} \right), \label{eq:15}
\end{eqnarray}
where ${\mathds O}$ and ${\mathds 1}$ denote the $(n-1) \times(n-1)$
null matrix and identity matrix. As a consequence, the dynamical
equations (\ref{eq:5}) take the form (\ref{eq:1}).

It is worth noting that while in classical mechanics phase space
coordinates correspond to observables, in quantum mechanics only
half of the phase space coordinates correspond to observables.
Specifically, if we write ${\hat\Pi}_i= |E_i\rangle\langle E_i|$ for
the observable corresponding to the projection operator onto the
$i$th normalised energy eigenstate, then
$p_i=\langle\psi|{\hat\Pi}_i |\psi\rangle$. Therefore the
coordinates $\{p_i\}_{i=1,2, \ldots,n-1}$ constitute a commuting
family of observables. The conjugate variables
$\{q_i\}_{i=1,2,\ldots,n-1}$, correspond to the relative phases, and
do not represent observables in the conventional sense.

\textit{Example 1}. We now apply the formalism to specific examples.
The first example is a system consisting of a pair of
spin-$\frac{1}{2}$ particles. For a generic Hamiltonian, we shall
impose the constraint that under the dynamics the initial state of
the system remains on a quadratic surface ${\mathcal Q}= {\mathcal
P}^1 \times {\mathcal P}^1 \subset \Gamma$ that contains the energy
eigenstates. Such a constraint implies that the quantum state can be
represented as a product state with respect to some choice of basis
elements. The Hilbert space is four dimensional and a generic state
can be expressed in the form
\begin{eqnarray}
\fl |\psi\rangle = \surd{p_1}\re^{-{\rm i}q_1}|E_1\rangle +
\surd{p_2} \re^{-{\rm i}q_2}|E_2\rangle + \surd{p_3} \re^{-{\rm
i}q_3} |E_3\rangle + (1-p_1-p_2-p_3)^{1/2}|E_4\rangle.
\label{eq:16}
\end{eqnarray}
The space of pure states is the three-dimensional space ${\mathcal
P}^3$, in which sits the two-dimensional product space ${\mathcal
Q}$. The constraint for the state to remain on ${\mathcal Q}$ is
therefore of an algebraic type---that is, $|\psi \rangle$ must lie
on the algebraic subspace ${\mathcal Q}$. If we write
$\{\psi_i\}_{i=1,\ldots,4}$ for the coordinates of the Hilbert space
vector $|\psi\rangle=(\psi_1,\psi_2, \psi_3,\psi_4)$, then a
necessary and sufficient condition for $|\psi\rangle$ to lie on
${\mathcal Q}$ is $\psi_1 \psi_4 = \psi_2 \psi_3$~\cite{gqm}.
Expressing the real and the imaginary parts of this condition in
terms of the coordinates chosen in (\ref{eq:16}) we find that the
constraint equations are given by
\begin{eqnarray}
\begin{array}{l} \sqrt{p_1p_4}\,\cos q_1 - \sqrt{p_2 p_3}
\,\cos(q_2 + q_3) = 0 \\ \sqrt{p_1p_4}\,\sin q_1 - \sqrt{p_2 p_3} \,
\sin(q_2 + q_3) = 0, \end{array}  \label{eq:18}
\end{eqnarray}
where for brevity we have written $p_4= 1-p_1- p_2-p_3$. If we
divide the first equation in (\ref{eq:18}) by $\sqrt{p_2 p_3} \,\cos
q_1$ and the second by $\sqrt{p_2 p_3} \,\sin q_1$, and compare the
results, we find that the constraint equations can be made
separable:
\begin{eqnarray}
\begin{array}{l} \Phi^1 = q_1 - q_2 - q_3 \\ \Phi^2 =
p_1(1-p_1-p_2-p_3) - p_2 p_3. \end{array}   \label{eq:19}
\end{eqnarray}

Before we proceed to derive the constrained dynamical equations, we
address the following question: What is the condition on ${\hat H}$
that will ensure that under unitary evolution an initial state that
lies on ${\mathcal Q}$ remains on ${\mathcal Q}$? The answer is
obtained by substituting the solution (\ref{eq:4}) of the unitary
motion into the constraints (\ref{eq:19}). We thus obtain the
condition that $\omega_1=\omega_2+\omega_3$. Translated into the
eigenvalues of ${\hat H}$ this condition is $E_1-E_2= E_3-E_4$. It
follows that the trace-free part of the Hamiltonian must have the
eigenvalue structure $\{e_1,e_2,-e_2,-e_1 \}$.

We now turn to the general case for which $E_1-E_2\neq E_3-E_4$. As
a consequence of the separable decomposition (\ref{eq:19}), the
matrix $\omega^{\alpha\beta}$ and its inverse $\omega_{\alpha\beta}$
are remarkably simple in this example:
\begin{eqnarray}
\omega^{\alpha\beta}=\left( \begin{array}{cc} {\mathds O} & {\mathds
1} \\ -{\mathds 1} & {\mathds O} \end{array} \right) \quad {\rm and}
\quad  \omega_{\alpha\beta}=\left( \begin{array}{cc} {\mathds O} &
-{\mathds 1} \\ {\mathds 1} & {\mathds O} \end{array} \right) .
\label{eq:20}
\end{eqnarray}
It follows that
\begin{eqnarray}
\Lambda^{ab}=\left( \begin{array}{cc} {\mathds O} & {\mathds A} \\
-{\mathds A} & {\mathds O} \end{array} \right), \label{eq:21}
\end{eqnarray}
where, writing $p_4= 1-p_1- p_2-p_3$ as before, we have
\begin{eqnarray}
{\mathds A}=\left( \begin{array}{ccc} p_1-p_4 & p_4-p_1 & p_4-p_1 \\
p_1+p_3 & -p_1-p_3 & -p_1-p_3 \\ p_1+p_2 & -p_1-p_2 & -p_1-p_2
\end{array} \right). \label{eq:22}
\end{eqnarray}
Substituting these results into (\ref{eq:13}) we obtain the
following equations of motion:
\begin{eqnarray}
\begin{array}{l} \dot{q}_1 =
(\omega_1-\omega_2-\omega_3)(2p_1+p_2+p_3)+(\omega_2+\omega_3) \\
\dot{q}_2 = (\omega_1-\omega_2-\omega_3)(p_1 + p_3) + \omega_2 \\
\dot{q}_3 = (\omega_1-\omega_2-\omega_3)(p_1 + p_2) + \omega_3 \\
\dot{p}_1 = 0 \\
\dot{p}_2 = 0 \\
\dot{p}_3 = 0. \end{array} \label{eq:23}
\end{eqnarray}
It should be evident that if the condition $\omega_1=\omega_2+
\omega_3$ holds, then (\ref{eq:23}) reduces to the unitary case. It
is interesting to observe that in spite of the fact that the
evolution is no longer unitary we nevertheless have ${\dot p}_i=0$
and hence ${\dot q}_i={\rm constant}$ for $i=1,2,3$. In other words,
the evolution is `quasi-unitary'.

\begin{figure}
\begin{center}\vspace{-0.0cm}
  \includegraphics[scale=0.45]{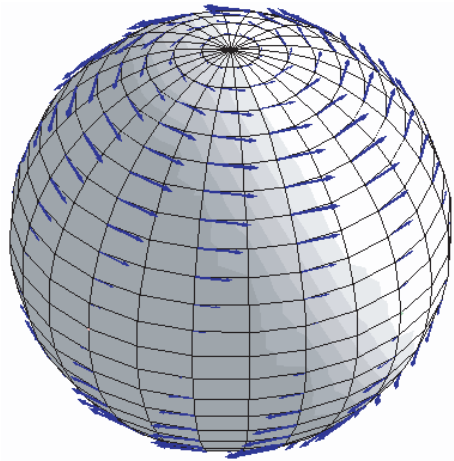}
  \vspace{-0.0cm}
  \caption{A `snapshot' of the vector field generated by
  $(\dot{\theta}_1, \dot{\phi}_1)$ in Example~1. The parameters are
  chosen to be $E_1=\frac{1}{2}$, $E_2=1$, $E_3=-2$,
  $E_4=\frac{1}{2}$, and $\theta_2=\half\pi$, thus giving
  $\dot{\theta}_1=0$ and $\dot{\phi}_1=\half-\sin^2\half\theta_1$.
  In this example, $\dot{\theta}_1>0$ in the northern hemisphere,
  $\dot{\theta}_1=0$ along the equator, and $\dot{\theta}_1<0$ in
  the southern hemisphere, resembling the nonlinear motion
  considered by Mielnik \cite{Mielnik2}.
  \label{fig:1}
  }
\end{center}
\end{figure}

To gain further insight into the dynamics generated by (\ref{eq:23})
we note that ${\mathcal Q}$ is the product of two Bloch spheres. Any
motion on ${\mathcal Q}$ thus corresponds to a pair of coupled
trajectories on these spheres. In terms of the usual spherical
coordinates $(\{\theta_i\},\{\phi_i\})_{i=1,2}$ a point on
${\mathcal Q}$ can be written in the form $|\psi_1\rangle|\psi_2
\rangle$, where
\begin{eqnarray}
|\psi_i \rangle=\cos\half\theta_i|\!\!\uparrow\rangle +
\sin\half\theta_i \,\re^{{\rm i} \phi_i}|\!\!\downarrow\rangle.
\end{eqnarray}
The idea of visualising the motions on the spheres is to express the
phase-space coordinates $(\{q_i\}, \{p_i\})$ in terms of the
spherical coordinates $(\{\theta_i\},\{\phi_i\})$. For this, we must
specify a Hamiltonian so that we can establish the relation between
the energy eigenstates and the four chosen basis states
$|\!\!\uparrow\uparrow\rangle, |\!\!\uparrow\downarrow\rangle,
|\!\!\downarrow\uparrow\rangle, |\!\!\downarrow\downarrow\rangle$ on
${\mathcal Q}$. For example, suppose that the Hamiltonian takes the
form
\begin{eqnarray}
{\hat H}=-J{\hat{\boldsymbol\sigma}}_1\otimes
{\hat{\boldsymbol\sigma}}_2 -B (\hat{\sigma}^z_{1} \otimes{\mathds
1}_2+{\mathds 1}_1 \otimes \hat{\sigma}^z_{2}), \label{eq:24}
\end{eqnarray}
which is a Heisenberg-type spin-spin interaction with strength $J$
and an external $z$-field with strength $B$. The eigenstates of this
Hamiltonian are given by the spin-$0$ singlet state and the spin-$1$
triplet states. By comparing the coefficients of
$|\psi_1\rangle|\psi_2 \rangle$ with (\ref{eq:16}) one can express
$(\{q_i\},\{p_i\})$ in terms of $(\{\theta_i\}, \{\phi_i\})$. We
invert the resulting relations to obtain
\begin{eqnarray}
\begin{array}{ccl} \theta_{1,2} &=& \sin^{-1}\sqrt{p_2 +
p_3 - 2\sqrt{p_1p_4}} \pm \cos^{-1}\sqrt{p_1 + p_4 -
\sqrt{p_1p_4}} \\ \phi_{1,2} &=& -\frac{1}{2}\left(q_1 \pm
\cos^{-1}\left(\half(p_3 - p_2)/\sqrt{p_1p_4} \right) \right),
\end{array}  \label{eq:25}
\end{eqnarray}
where $(\theta_1,\phi_1)$ corresponds to the `$+$' sign and
$(\theta_2, \phi_2)$ corresponds to the `$-$' sign in the right side
of (\ref{eq:25}). From (\ref{eq:23}) and (\ref{eq:25}) we deduce the
equations of motion in terms of the spherical variables, with the
results:
\begin{eqnarray}
\begin{array}{l} \dot{\theta}_1 = \dot{\theta}_2 = 0 \\
\dot{\phi}_1 = \dot{\phi}_2 = - \half(\omega_1 - \omega_2 -
\omega_3) (\sin^2 \half \theta_1 + \sin^2\half\theta_2) -\half
(\omega_2 + \omega_3). \end{array}
\end{eqnarray}
We thus find that for a given initial state the dynamical
trajectories are given by a pair of latitudinal circles on the
respective Bloch spheres. An example of a field plot for one of the
Bloch spheres is shown in Figure~\ref{fig:1}.

\textit{Example 2}. We consider a system of three spin-$\frac{1}{2}$
particles, and impose the restriction that the unitary evolution is
constrained to lie on a product space
$\mathcal{R}=\mathcal{P}^1\times \mathcal{P}^1 \times \mathcal{P}^1$
that contains the energy eigenstates. In this case, a necessary and
sufficient condition for the state $|\psi\rangle$ to lie on
$\mathcal{R}$ is that the components of the state vector
simultaneously satisfy two quadratic equations~\cite{bgh}. Expressed
in terms of the homogeneous coordinates $\{\psi_i\}_{i=1,2,
\ldots,8}$ of $|\psi\rangle$ we find that the relevant constraints
are given by four complex equations: $\psi_1\psi_7=\psi_3\psi_4$,
$\psi_2\psi_8=\psi_5\psi_6$, $\psi_1\psi_8=\psi_2\psi_7$, and
$\psi_3\psi_6=\psi_4\psi_5$. Taking the real and the imaginary parts
of these equations and expressing the results in terms of
$(\{q_i\},\{p_i\})$ we obtain the following eight constraints:
\begin{eqnarray}
\begin{array}{l} \Phi^1 = q_2 - q_5 - q_6 \\
\Phi^2 = q_1 + q_7 - q_3 - q_4 \\
\Phi^3 = q_1 - q_2 - q_7 \\
\Phi^4 =q_3+q_6-q_4-q_5 \\
\Phi^5 = p_2 p_8 - p_5p_6 \\
\Phi^6 = p_1 p_7 - p_3 p_4 \\
\Phi^7 = p_1 p_8 - p_2 p_7 \\
\Phi^8 = p_3 p_6 - p_4 p_5, \end{array}  \label{eq:26}
\end{eqnarray}
where we have written $p_8=1-\sum_{i=1}^7 p_i$. Remarkably, the
constraint equations for the variables $\{q_i\}$ and $\{p_i\}$
decouple. We can also read off from (\ref{eq:26}) the condition for
the unitary motion to lie on $\mathcal{R}$:
\begin{eqnarray}
\begin{array}{l}
\omega_1 = \omega_2 + \omega_7 \\
\omega_2 =\omega_5+\omega_6 \\
\omega_1 + \omega_7= \omega_3 + \omega_4 \\
\omega_3 + \omega_6 = \omega_4 + \omega_5. \end{array}
\end{eqnarray}
It follows that the eigenvalues of the
trace-free part of the Hamiltonian must take the form $\{e_1, e_2,
e_3, e_4, -e_4, -e_3, -e_2, -e_1\}$.

If ${\hat H}$ does not have this property, then the constraints
(\ref{eq:26}) become nontrivial. Nevertheless, owing to the fact
that they decouple it is straightforward to verify that the
equations of motion are given by $\dot{q}_i=f_i(\{p_i\})$ and
$\dot{p}_i=0$ for $i=1,\ldots,7$, where $f_i(\{p_i\})$ are
elementary functions of the variables $\{p_i\}$. It follows that the
dynamical evolution is quasi-unitary in that the amplitudes
$\{p_i\}$ remain constant and the relative phases $\{p_i\}$ evolve
linearly in time.

\textit{Example 3}. Let us turn to a different example. We consider
again a pair of spin-$\frac{1}{2}$ particles and impose the
condition that an initially disentangled quantum state remains
disentangled under the dynamics. This constraint can be expressed
algebraically by requiring that the motion is confined to the
special quadric $\mathcal{Q'}$ that corresponds to disentangled spin
states. A generic state is still given by (\ref{eq:16}) but it is
now no longer the case that all the energy eigenstates $|E_i\rangle$
lie on $\mathcal{Q}'$. In other words, the constraint surface is no
longer given by $\psi_1\psi_4=\psi_2 \psi_3$ in the energy basis.

\begin{figure}
\begin{center}\vspace{-0.0cm}
  \includegraphics[scale=0.45]{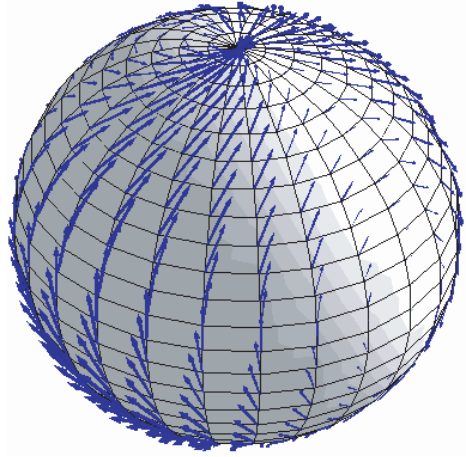}
  \vspace{-0.0cm}
  \caption{A snapshot of the vector field generated by
  $(\dot{\theta}_1,\dot{\phi}_1)$ in Example 3. The parameters are
  chosen as $E_1=\frac{1}{2}$, $E_2=1$, $E_3=-2$, $E_4=\frac{1}{2}$,
  and $\theta_2=\phi_2=\half\pi$. This choice gives the vector field
  $\dot{\theta}_1= \cos \phi_1 \left(\frac{1}{2}\cos\theta_1
  -3 \right)$ and $\dot{\phi}_1=\frac{1}{4}\left(9\cos\theta_1-
  \sin\phi_1 \cos^2\theta_1/\sin\theta_1\right)$.
  \label{fig:2}
  }
\end{center}
\end{figure}

For the specification of the constraint equation associated with
disentangled states, we must specify the Hamiltonian. We consider
the Heisenberg-type spin-spin interaction system given by
(\ref{eq:24}). The corresponding energy eigenstates include the
total spin-$0$ singlet state and the total spin-$1$ triplet states.
We then change basis by substituting the energy eigenstates in
(\ref{eq:16}) with the singlet and triplet spin states such that we
can express our general state in terms of the four disentangled
basis states $|\!\!\uparrow\uparrow\rangle,
|\!\!\uparrow\downarrow\rangle, |\!\!\downarrow\uparrow\rangle,
|\!\!\downarrow\downarrow\rangle$ of $\mathcal{Q}'$. We find
\begin{eqnarray}
|\psi\rangle & = & {\surd p_4}|\!\!\uparrow\uparrow\rangle +
\frac{1}{{\surd 2}}({\surd p_2} \re^{-{\rm i} q_2} - {\surd p_3}
\re^{-{\rm i} q_3})|\!\!\uparrow\downarrow\rangle \nonumber \\ {} &
{} & + \frac{1}{{\surd 2}}({\surd p_2} \re^{-{\rm i} q_2} + {\surd
p_3} \re^{-{\rm i} q_3})|\!\!\downarrow\uparrow\rangle + {\surd p_1}
\re^{-{\rm i} q_1}|\!\!\downarrow\downarrow \rangle. \label{eq:37}
\end{eqnarray}
Since all the basis states used in (\ref{eq:37}) lie on
$\mathcal{Q}'$, now we are able to use the condition
$\psi_1\psi_4=\psi_2\psi_3$ to constrain the motion on
$\mathcal{Q}'$:
\begin{eqnarray}
\sqrt{p_1 p_4} \re^{-{\rm i} q_1} = \half(p_2 \re^{-2{\rm i} q_2} -
p_3 \re^{-2{\rm i} q_3}). \label{eq:28}
\end{eqnarray}
Taking the real and imaginary part of (\ref{eq:28}) we find that the
two constraints required for the motion to remain on $Q'$ are
\begin{eqnarray}
\begin{array}{l} \Phi_1  = 2\sqrt{p_1 p_4} - p_2\cos(2q_2 -
q_1) + p_3 \cos(2q_3-q_1) = 0 \\ \Phi_2  = p_2 \sin(2q_2 - q_1) -
p_3 \sin(2q_3-q_1)  = 0. \end{array}  \label{eq:29}
\end{eqnarray}
Unlike the previous examples, these constraints are not separable,
hence we can no longer expect to find the quasi-unitary motion seen
in the previous cases. Substituting the constraints (\ref{eq:29})
into (\ref{eq:10}) and following the procedures, we find that the
resulting equations of motion are given by
\begin{eqnarray}
\fl \begin{array}{l}
\dot{p}_1 = 2 p_2 p_3 (\omega_1 - \omega_3) \sin(2(q_2-q_3)) \\
\dot{p}_2 = - 2 p_2 p_3 (\omega_1 - 2\omega_3) \sin(2(q_2-q_3)) \\
\dot{p}_3 =  2 p_2 p_3 (\omega_1 - 2\omega_2) \sin(2(q_2-q_3)) \\
\dot{q}_1 =  2 p_3(1-2p_1 - p_2 -p_3)(\omega_2-\omega_3) \cos(2q_3 -
q_1)/\sqrt{p_1p_4} \\
\qquad + (\omega_1 - 2\omega_2)(2p_1 + p_2 +p_3) + 2\omega_2 \\
\dot{q}_2  =  2 p_1 p_3 (\omega_3 - \omega_2) \cos(2q_3 -
q_1)/\sqrt{p_1 p_4} + (\omega_1 - 2\omega_2)(2p_1 + p_2) -
p_3(\omega_1-2\omega_3) + 2\omega_2 \\ \dot{q}_3  = 2 p_1 p_3
(\omega_3 - \omega_2) \cos(2q_3 - q_1)/\sqrt{p_1 p_4} \\ \qquad +
(\omega_1 - 2\omega_2)(2p_1 - p_2\cos(2(q_2-q_3))) + p_3
(\omega_1-2\omega_3) + 2\omega_3. \\
\end{array} \label{eq:400}
\end{eqnarray}
It is no longer the case that $\dot{p}_i=0$. Therefore, we have six
coupled nonlinear differential equations describing the motion. We
can visualise the motion on the product of two Bloch spheres by
converting (\ref{eq:400}) into spherical coordinates:
\begin{eqnarray}
\fl  \begin{array}{l}
\dot{\theta}_1 = \sin(\phi_1 - \phi_2)
\sin\theta_2 [ (\omega_1 - \omega_2) \cos\theta_1 + \omega_2 -
\omega_3]\\
\dot{\theta}_2  =  \sin(\phi_1 - \phi_2) \sin\theta_1
[(\omega_2 - \omega_1)\cos\theta_2 - \omega_2 + \omega_3] \\
\dot{\phi}_1 =  \frac{1}{2} [-\omega_1 + (\omega_2 -
\frac{\omega_1}{2})\cos\theta_2 + (\frac{3}{2} \omega_1 - \omega_2
-2\omega_3)\cos\theta_1 + [\cos(\phi_1-\phi_2)/(\sin \theta_1
\sin\theta_2)] \\ \qquad \quad \times \left.\left(2
(\omega_3-\omega_2) \sin^2\theta_1 \cos\theta_2 + (\omega_1-
\omega_2) (\cos^2\theta_1 - \cos^2\theta_2) \right)\right] \\
\dot{\phi}_2  =  \frac{1}{2} [-\omega_1 + (\omega_2 -
\frac{\omega_1}{2})\cos\theta_1 + (\frac{3}{2} \omega_1 - \omega_2
-2\omega_3)\cos\theta_2 + [\cos(\phi_1-\phi_2)/(\sin\theta_1
\sin\theta_2)] \\ \qquad \quad \times \left.
\left(2(\omega_3-\omega_2)\cos\theta_1 \sin^2\theta_2 - (\omega_1
-\omega_2)(\cos^2\theta_1 - \cos^2\theta_2) \right) \right].
\end{array}
\end{eqnarray}
An example of the resulting field plot arising from these equations
is shown in Figure~\ref{fig:2}, indicating the nontrivial nature of
the dynamics.

In summary, we see that constrained quantum motions of the Dirac
type can be treated straightforwardly by means of the prescription
described above. The resulting dynamical equations are in general
quite intricate, and typically require numerical analysis.

\vspace{0.5cm}
\begin{footnotesize}
The authors thank D.~Holm and M.~Parry for stimulating discussions.
This work was carried out in part while DCB and LPH were visitors at
the Perimeter Institute for Theoretical Physics, Waterloo, Ontario.
\end{footnotesize}
\vspace{0.5cm}



\end{document}